\documentclass[conference]{IEEEtran}
\usepackage{epsfig,cite}
\usepackage{graphicx}
\usepackage{color}
\usepackage{amssymb,amsmath,mathtools, mathrsfs}
\usepackage{enumitem}
\usepackage{graphicx}
\usepackage{epstopdf}
\definecolor{blue}{rgb}{0,0,1}
\definecolor{darkgreen}{rgb}{0,.5,0}
\definecolor{darkred}{rgb}{.5,0,0}

\newtheorem{theorem}{Theorem}

\newtheorem{lemma}{Lemma}
\newtheorem{rem}{Remark}

\newtheorem{definition}{Definition}

\def\levy{L\'evy }

\newcommand{\realSet}{\mathcal{R}}

\def\LevyDist{{\mathscr{L}}}
\def\Pr{{\mathrm{Pr}}}

\DeclareMathOperator\erfc{erfc}

\IEEEoverridecommandlockouts

\begin{document}

\title{On the Capacity of Diffusion-Based \\Molecular Timing Channels}

\author{\IEEEauthorblockN{Nariman Farsad\IEEEauthorrefmark{1}, Yonathan~Murin\IEEEauthorrefmark{1}, Andrew~Eckford\IEEEauthorrefmark{2}, and Andrea Goldsmith\IEEEauthorrefmark{1}}\IEEEauthorblockA{\IEEEauthorrefmark{1}Electrical Engineering, Stanford University, USA } \IEEEauthorblockA{\IEEEauthorrefmark{2}Electrical Engineering \& Computer Science, York University, Canada}
\vspace{-0.7cm}
}
%
%
\maketitle

\begin{abstract}
	This work introduces capacity limits for molecular timing (MT) channels, where information is modulated on the release timing of small information particles, and decoded from the time of arrival at the receiver. It is shown that the random time of arrival can be represented as an additive noise channel, and for the diffusion-based MT (DBMT) channel, this noise is distributed according to the \levy distribution. Lower and upper bounds on the capacity of the DBMT channel are derived for the case where the delay associated with the propagation of information particles in the channel is finite. These bounds are also shown to be tight. 
\end{abstract}


\section{Introduction}

Molecular communication is a new and emerging field where small particles such as molecules are used to transfer information \cite{far16ST}. Information can be modulated on different properties of these particles such as their concentration, 
the type, 
the number, 
or the time of release. 
 Moreover, different techniques can be used to transfer the particles from the transmitter to the receiver including: diffusion, 
active transport, 
bacteria, 
and flow. 
To show the feasibility of molecular communication, several experimental systems that are capable of transmitting short messages at low bit rates have been developed in recent years \cite{far13}.
Yet, despite all these advancements, there are still many open problems in the field, especially from an information theoretic perspective. For example, the channel capacity of many different molecular communication systems is still unknown \cite{far16ST}, particularly those with
indistinguishable molecules \cite{rose14}. 

In this work, we consider molecular communication systems where information is modulated on the {\em time of release of the information particles}. In biology, time of release may be used in the brain at the synaptic cleft, where two chemical synapses communicate over a chemical channel \cite{bor99}. 
A common assumption, which is accurate for many sensors, is that the particle is detected and then removed from the environment as part of the detection process. Thus, the random delay until the particle first arrives at the receiver can be represented as an additive noise term. For example, for diffusion-based channels, the random first time of arrival is L\'evy-distributed \cite{karatzas-shreve}. 

Although there are similarities between the timing channel considered in this work and the timing channel considered in \cite{ana96}, which studied the transmission of bits through queues, the problem formulation and the noise models are fundamentally different. In \cite{ana96}, the queue induces an order on the channel output (i.e. arrival times), namely, the first arrival time corresponds to the first channel use, the second arrival corresponds to second channel use, and so on. 
On the other hand, in molecular channels with {\em indistinguishable} particles, {\em order may not be preserved}, as was observed in \cite{rose14}. 
Regarding the differences in the noise models, we note that in \cite{ana96} the random delay is governed by the queue's service distribution, while in molecular communication the random characteristics is associated with the transport of information particles in molecular channels.


Some of the previous works on molecular timing channels focused on the additive inverse Gaussian noise (AIGN) channel, which features a positive drift from the transmitter to the receiver \cite{sri12,cha12,eck12,li14}. In this case, the first time of arrival over a one-dimensional space follows the inverse Gaussian distribution, giving the channel its name. In these works, the upper and lower bounds on the maximal mutual information between the AIGN channel input and output, denoted in this work by {\em capacity per channel use}, were provided for different input and output constraints.
One of the main unresolved issues in these works is the problem of ordering when information particles from consecutive channel uses may arrive out of order (i.e. during other channel use intervals).  Thus, it is not clear from \cite{sri12, cha12,eck12,li14} how information can be transmitted sequentially, and what is the associated capacity in {\em bits per second}. 

To deal with this challenge, in the current work we make two assumptions. 
First, we assume that there is a finite time interval, called the {\em symbol interval}, over which the transmitter can release the information particles (the message to be transmitted is encoded in this time). 
Second, we assume that the information particles have a finite lifespan, which we call the {\em particle's lifetime}. The underlying assumption is that the particles are dissipated immediately after this time interval.
We note that this assumption can be incorporated into a system by using enzymes or other chemicals that degrade the particles \cite{guo15}; as long as the particle's lifetime is less than infinity, our results and analysis hold. 
Using these assumptions, a {\em single channel use interval} is the sum of the symbol interval and the particle's lifetime. Information particles arrive during the same channel use in which they were released, or they dissipate over this interval and hence never arrive. 
These assumptions enforce an ordering in which particles arrive in the same order in which they are transmitted, which allows identical and independent consecutive channel uses. 
In this work, we introduce this channel as the {\em molecular timing} (MT) channel, which can be used with any propagation mechanism as long as the particles follow independent paths, and have a finite lifetime and symbol interval. 
Using this formulation, we formally define the capacity of the MT channel in bits per second. 
We then apply this definition to the diffusion-based MT (DBMT) channel, where the particles diffuse without a drift from the transmitter to the receiver, and derive an upper and a lower bound on the capacity of this channel. These are the {\em first bounds} on the capacity of diffusion-based molecular timing channels.
Numerical evaluations indicate that these bounds can be tight. 

The rest of this paper is organized as follows. The system model and the problem formulation are presented in Section \ref{sec:model}. The capacity of the DBMT channel is studied in Section \ref{sec:singlePart}, while lower and upper bounds on this capacity are derived in Section \ref{sec:capBounds}. 
Numerical evaluations are presented in Section \ref{sec:Numerical}, and concluding remarks are provided in Section \ref{sec:concl}.

{\bf {\slshape Notation}:} 
We denote random variables (RVs) with upper case letters, $X$, and their realizations with the corresponding lower case letters, e.g., $x$. 
Sets are denoted by calligraphic letters, e.g., $\mathcal{A}$, where $\realSet$ is the set of real numbers. 
$f_{X}(x)$ is used to denote the probability density function (PDF) of a continuous RV $X$ on $\realSet$, and $F_{X}(x)$ its cumulative distribution function (CDF). 
$\erfc\left( \cdot \right)$ is used to denote the complementary error function and $\log (\cdot)$ is used to denote the logarithm with basis 2. 
Finally, $h(\cdot)$, $I(\cdot;\cdot)$, and $X \leftrightarrow Y \leftrightarrow Z$ are used to denote differential entropy, mutual information, and a Markov chain, respectively \cite{kim-elgamal-book}.

\vspace{-0.2cm}
\section{System Model and Problem Formulation}
\label{sec:model}

\vspace{-0.1cm}
\subsection{The Molecular Timing Channel} \label{subsec:ALN}

\vspace{-0.1cm}
We consider a molecular communication channel in which information is modulated on the time of release  of the information particles. 
The information particles themselves are assumed to be {\em identical and indistinguishable} at the receiver. Therefore, the receiver can only use the time of arrival to decode the intended message.
The information particles propagate from the transmitter to the receiver through some random propagation mechanism (e.g. diffusion). To develop our model, we make the following assumptions about the~system:

\begin{enumerate}[label = {\bf A{\arabic*}})]
	\item \label{assmp:perfectTxRx}
	The transmitter and receiver are perfectly synchronized in time. The transmitter perfectly controls the release time of the particles, while the receiver perfectly measures the arrival times. 
	
	\item \label{assmp:Arrival}
	An information particle which arrives at the receiver is absorbed and removed from the propagation medium.
	
	\item \label{assmp:indep}
	All information particles propagate independently of each other, and their trajectories are random according to an independent and identically distributed (i.i.d.) process. 
\end{enumerate} 
\noindent Note that these assumptions have been traditionally considered in all previous works \cite{sri12, cha12, eck12, li14, rose14} to make the models tractable.

Next, we formally define the channel. 
Let $T_{x,k} \mspace{-3mu} \in \mspace{-3mu} \realSet^+, k \mspace{-3mu} = \mspace{-3mu}1,2,\dots,\ell$, denote the time of the $k$th transmission. At $T_{x,k}$, a single information particle is released into the medium by the transmitter. The transmitted information is encoded in the sequence of times $\{T_{x,k}\}_{k=1}^\ell$, where $\{T_{x,k}\}_{k=1}^\ell$ are assumed to be independent of the random propagation times of {\em each} of the information particles.
Let $T_{y,k}$ denote the time of arrival of the information particle transmitted at $\{T_{x,k}\}$, thus, it follows that $T_{y,k} \mspace{-3mu} \geq \mspace{-3mu} T_{x,k}$, which leads to the following additive noise~channel model:
\vspace{-0.15cm}
\begin{align}
	\label{eq:levyChan}	
	T_{y,k} = T_{x,k} + T_{n,k}, 
\end{align}

\vspace{-0.1cm}
\noindent where $T_{n,k}$, is a random noise term representing the propagation time of the particle transmitted at $T_{x,k}$.

One of the properties of the channel \eqref{eq:levyChan} is that order is not preserved, namely, the arrival order may differ from the transmission order, which results in a channel with memory. 
To resolve this issue, we make two assumptions. 
First, we assume that in each channel use interval the transmitter releases the information particle within its finite {\em symbol interval}. 
Second, we assume that information particles have a finite lifetime, i.e., they dissipate immediately after this finite interval, denoted by the {\em particle's lifetime}. 
By setting the channel use interval to be a concatenation of the symbol interval and the particle's lifetime, we ensure that order is preserved and obtain a memoryless channel.

Let $\tau_x < \infty$ be the symbol interval, and  $\tau_n < \infty$ be the particle's lifetime, i.e., each time slot is of length $\tau_x+\tau_n$. The above two assumptions can now be formally stated as:
\begin{enumerate}[label = {\bf A{\arabic*}}), resume]
	\item	
	The release times obey:
	\vspace{-0.15cm}
	\begin{align*}
	(k-1)\cdot (\tau_x + \tau_n) \le T_{x,k} \leq (k-1)\cdot (\tau_x + \tau_n) + \tau_x.
	\end{align*}		
	
	\vspace{-0.15cm}
	\item \label{assmp:lifetime}
	The information particles dissipate and are never received if $T_{n,k} \geq \tau_n$. \label{asmp:limNoise}
	
\end{enumerate}

\noindent The first assumption can be justified by noting that the transmitter can choose its release interval, while the second assumption can be justified by designing the system such that information particles are degraded in the environment after a finite time (e.g. using chemical reactions) \cite{guo15}. 
The resulting channel, which we call the {\em molecular timing (MT) channel}, is given by:
\vspace{-0.15cm}
\begin{align}
\label{eq:TCALNchan}
Y_k = \begin{cases} T_{y,k} = T_{x,k} + T_{n,k}, & T_{n,k} \leq \tau_n \\ \phi, & T_{n,k} > \tau_n \end{cases},
\end{align}

\begin{figure}
	\begin{center}
		\includegraphics[width=0.7\columnwidth,keepaspectratio]{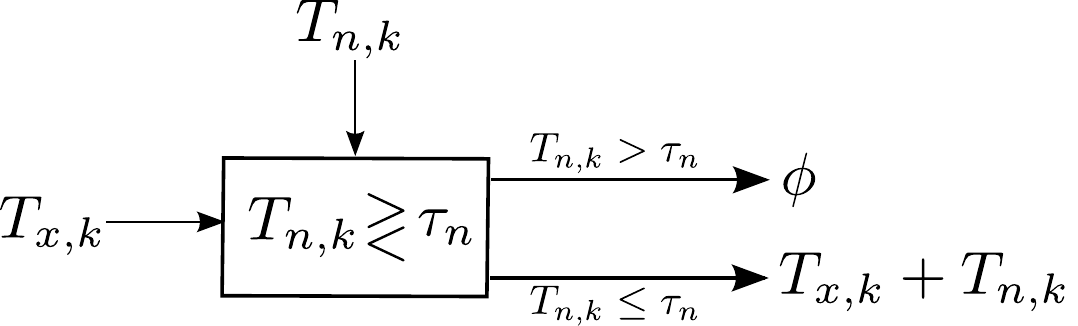}
	\end{center}
	\vspace{-0.5cm}
	\caption{\label{fig:chanModel} The MT channel in \eqref{eq:TCALNchan}. The channel input is $T_{x,k}$, while the channel output depends on the condition $T_{n,k} \gtrless \tau_n$.  }
	\vspace{-0.5cm}
\end{figure}

\vspace{-0.10cm}
\noindent where $\phi$ is the empty symbol, a symbol that indicates no particle arrived. This channel is depicted in Fig.~\ref{fig:chanModel}. Next, we formally define the capacity of the MT channel.

\vspace{-0.15cm}
\subsection{Capacity Formulation for the MT Channel} \label{subsec:ProbDef}

Let $\mathcal{A}_k \mspace{-2mu} \triangleq \mspace{-2mu} [(k \mspace{-2mu} - \mspace{-2mu} 1)\cdot (\tau_x \mspace{-2mu} + \mspace{-2mu} \tau_n), (k \mspace{-2mu} - \mspace{-2mu} 1)\cdot (\tau_x \mspace{-2mu} + \mspace{-2mu} \tau_n) \mspace{-2mu} + \mspace{-2mu} \tau_x]$ and $\mathcal{B}_k \mspace{-2mu} \triangleq \mspace{-2mu} \left\{ [(k \mspace{-2mu} - \mspace{-2mu} 1)\cdot (\tau_x \mspace{-2mu} + \mspace{-2mu} \tau_n), k\cdot (\tau_x \mspace{-2mu} + \mspace{-2mu} \tau_n)] \cup \phi \right\}$ for $k \mspace{-2mu} = \mspace{-2mu} 1,2,\dots,\ell$.
We now define a code for the MT channel~\eqref{eq:TCALNchan} as~follows:
\begin{definition}[Code] \label{def:Encoding}
	A $(\ell, \mathsf{R}, \tau_x, \tau_n)$ code for the MT channel \eqref{eq:TCALNchan}, with code length $\ell$ and code rate $\mathsf{R}$, consists of a message set $\mathcal{W} = \{1,2,\dots,2^{\ell(\tau_x+\tau_n)\mathsf{R}} \}$, an encoder function $\varphi^{(\ell)}: \mathcal{W} \mapsto \mathcal{A}_1 \times \mathcal{A}_2 \times \dots \times \mathcal{A}_{\ell}$, and a decoder function  $\nu^{(\ell)}: \mathcal{B}_1 \times \mathcal{B}_2 \times \dots \times \mathcal{B}_{\ell} \mapsto \mathcal{W}$.
\end{definition}

\begin{rem}
	Observe that since we consider a timing channel, similarly to \cite{ana96}, the codebook size is a function of $\tau_x + \tau_n$, and $\ell(\tau_x + \tau_n)$ is the maximal time that it takes to transmit a message using a $(\ell, \mathsf{R}, \tau_x, \tau_n)$ code. 
	Furthermore, note that the above encoder maps the message $W \in \mathcal{W}$ into $\ell$ time indices, $T_{x,k},k=1,2,\dots,\ell$, where $T_{x,k} \in \mathcal{A}_k$, while the decoder decodes the transmitted message using the $\ell$ channel outputs $\{Y_k\}_{k=1}^\ell, Y_k \in \mathcal{B}_k$. We emphasize that this construction creates an {\em ordering} of the different arrivals, which leads to $\ell$ identical and independent channels.
	Finally, we note that this construction was not used in \cite{ana96} since when transmitting bits through queues the channel itself forces an ordering.
\end{rem}

\begin{figure}[!t]
	\begin{center}
		\includegraphics[width=\columnwidth,keepaspectratio]{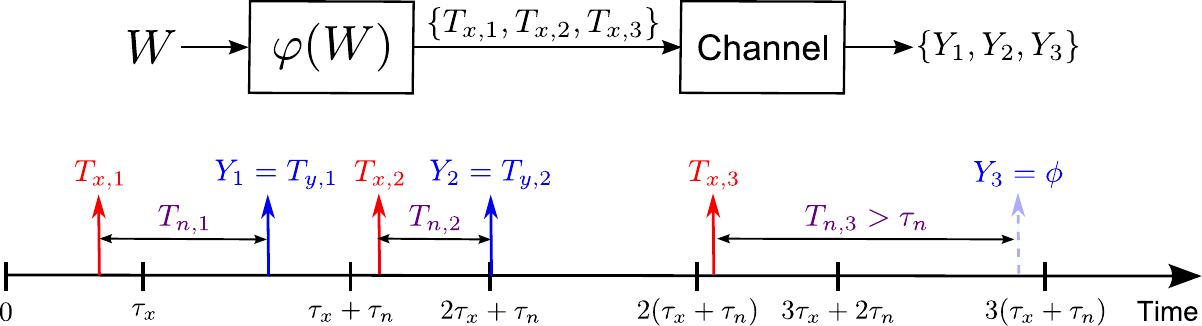}
	\end{center}
	\vspace{-0.3cm}
	\caption{\label{fig:Encoding} Illustration of the encoding procedure of Definition \ref{def:Encoding} for $\ell=3$. Red pulses correspond to transmission times, while blue pulses correspond to arrival times at the receiver.}
	\vspace{-0.25cm}
\end{figure}

\begin{rem}
The encoding and transmission are illustrated in Fig. \ref{fig:Encoding} for $\ell=3$. The encoder produces three release times $\{T_{x,1}, T_{x,2}, T_{x,3}\}$ which obey $T_{x,k} \in \mathcal{A}_k, k=1,2,3$. In each time index a single particle is released to the channel which adds a random delay according to \eqref{eq:TCALNchan}. The channel outputs are denoted by $\{Y_1,Y_2,Y_3\}$. It can be observed that while $Y_1 \mspace{-3mu} = \mspace{-3mu} T_{y,1} \mspace{-3mu} = \mspace{-3mu} T_{x,1} \mspace{-3mu} + \mspace{-3mu} T_{n,1}$ and $Y_2 \mspace{-3mu} = \mspace{-3mu} T_{y,2} \mspace{-3mu} = \mspace{-3mu} T_{x,2} \mspace{-3mu} + \mspace{-3mu} T_{n,2}$, $Y_3 \mspace{-3mu} = \mspace{-3mu} \phi$ since $T_{n,3} \mspace{-3mu} > \mspace{-3mu} \tau_n$ and therefore the third particle does not arrive.
\end{rem}

\begin{definition}[Probability of Error]
	The average probability of error of an $(\ell, \mathsf{R}, \tau_x, \tau_n)$ code is defined as: 
	\vspace{-0.15cm}
	\begin{align*}
	P_e^{(\ell)} \triangleq \Pr \left\{ \nu(Y_1, Y_2, \dots, Y_{\ell}) \neq W \right\},
	\end{align*}
	
	\vspace{-0.15cm}
	\noindent where the message $W$ is selected uniformly from $\mathcal{W}$.
	
\end{definition}

\begin{definition}[Achievable Rate and Capacity]
	A rate $\mathsf{R}$ is called achievable if for any $\epsilon > 0$ and $\delta > 0$ there exists some blocklength $\ell_0(\epsilon, \delta)$ such that for every $\ell> \ell_0(\epsilon, \delta)$ there exits an $(\ell, \mathsf{R}-\delta, \tau_x, \tau_n)$ code with $P_e^{(\ell)} < \epsilon$. The capacity $\mathsf{C}$ is defined to be the supremum of all achievable rates.
\end{definition}

\begin{rem}
	Note that even though we consider a timing channel, we define the capacity in terms of bits per time unit \cite[Definition 2]{ana96}. This is in contrast to the works \cite{sri12, cha12,eck12,li14} which defined the capacity as the maximal number of bits which can be conveyed through the channel {\em per channel use}.
\end{rem}

Note that this definition of capacity $\mathsf{C}$ for the MT channels is fairly general and can be applied to different propagation mechanism as long as Assumptions \ref{assmp:perfectTxRx}--\ref{assmp:lifetime} are not violated. 
Next, we focus on characterizing the capacity of MT channels with diffusion-based propagation.

\vspace{-0.1cm}
\subsection{The Diffusion-Based MT Channel}
In diffusion-based propagation, the released information particles follow a random Brownian path from the transmitter to the receiver. In this case, to specify the random additive noise term $T_{n,k}$ in \eqref{eq:TCALNchan}, we define a L\'evy-distributed RV as follows:
\begin{definition}[\levy Distribution] \label{def:levyRV}
	Let the RV $Z$ be a L\'evy-distributed RV with location parameter $\mu$ and scale parameter $c$. 
	Then, its PDF is given by
	\vspace{-0.15cm}
	\begin{align}
	\label{eqn:LevyPDF_0}
	f_Z(z)=
	\begin{cases}
	\sqrt{\frac{c}{2 \pi (z-\mu)^3}}\exp \left( -\frac{c}{2(z-\mu)} \right), & z>\mu \\
	0, & z\leq \mu
	\end{cases},
	\end{align} 
	
	\vspace{-0.15cm}
	\noindent and its CDF is given by
	\vspace{-0.15cm}
	\begin{align}
	\label{eqn:LevyCDF}
	F_Z(z) = \begin{cases} \erfc\left(\sqrt{\frac{c}{2(z-\mu)}}\right), & z>\mu \\ 0, & z\leq\mu \end{cases}.		
	\end{align}
\end{definition}

\vspace{-0.15cm}
\noindent Throughout the paper, we use the notation $Z \sim \LevyDist(\mu,c)$ to indicate a \levy RV with parameters $\mu$ and $c$.

Let $r$ denote the distance between the transmitter and the receiver, and $d$ denote the diffusion coefficient of the information particles in the propagation medium.  Following along the lines of the derivations in \cite[Sec. II]{sri12}, and using \cite[Sec. 2.6.A]{karatzas-shreve}, it can be shown that for 1-dimensional pure diffusion, the propagation time of each of the information particles follows a \levy distribution, and therefore the noise in \eqref{eq:TCALNchan} is distributed as $T_{n,k} \sim \LevyDist(0,c)$ with $c = \frac{r^2}{2d}$. 
In this case, we call the diffusion-based MT channel in \eqref{eq:TCALNchan} the {\em DBMT channel}.

\begin{rem}
	In \cite{yilmaz20143dChannelCF} it is shown that for an infinite,  three-dimensional homogeneous medium without flow, and a spherically absorbing receiver, the first arrival time follows a scaled \levy distribution. Thus, the results presented in this paper can be extended to 3-D space. 
\end{rem}

\section{The Capacity of the DBMT Channel} \label{sec:singlePart}

In this section we study the capacity of the DBMT channel.
Let $\mathcal{F}(\tau_x)$ denote the set of all PDFs $f_{T_x}(t_x)$ such that $F_{T_x}(t)=0$ for $t<0$ and $F_{T_x}(\tau_x)=1$.
The following theorem presents an expression for the capacity of the DBMT channel in~\eqref{eq:TCALNchan}. 
\begin{theorem} \label{thm:singlePartCap}
The capacity of the DBMT channel in \eqref{eq:TCALNchan} is given by:
\vspace{-0.15cm}
\begin{align}
\label{eq:TCALNchanCap}
 \mathsf{C}(\tau_n) \mspace{-3mu} =  \mspace{-3mu} \underset{\tau_x, \mathcal{F}(\tau_x)}{\max}  \frac{I(T_x;T_y|T_n<\tau_n)F_{T_n}(\tau_n)}{\tau_x + \tau_n}.
\end{align} 	
\end{theorem}

\vspace{-0.1cm}
\begin{IEEEproof}
	Using standard techniques, i.e., random coding \cite[Ch. 3.1.2]{kim-elgamal-book}, and information inequalities based on Fano's inequality \cite[Ch. 3.1.4]{kim-elgamal-book}, we show in \cite[Appendix A]{journalVersion} that the capacity of the channel \eqref{eq:TCALNchan}, in bits per second, is given by:
	\vspace{-0.15cm}
	\begin{align}
		\label{eq:capacityDef}
		\mathsf{C}(\tau_n) = \underset{\tau_x, \mathcal{F}(\tau_x)}{\max} \frac{I(T_x; Y)}{\tau_x + \tau_n}.
	\end{align}
	
	\vspace{-0.1cm}
	To evaluate \eqref{eq:capacityDef}, we first note that the channel \eqref{eq:TCALNchan} can be represented as two separate channels, where at each channel use only one is selected at random for transmission. 
This is illustrated in Fig. \ref{fig:chanModel}. Let $\Theta$ be a Bernoulli random variable that indicates which channel is selected at random with $\Theta=1$ if $T_n \leq \tau_n$, and $\Theta=0$ if $ T_n > \tau_n $.
Hence, $\Theta$ has a probability of success $p=F_{T_n}(\tau_n)$. 
	Since for each case the received symbol sets are disjoint, we have the Markov chain $T_x \leftrightarrow Y \leftrightarrow \Theta$. We next write:
	\vspace{-0.15cm}
	\begin{align}
		I(T_x;Y) &= I(T_x;Y,\Theta) \label{eqn:baseMI_markov} \\
				&=I(T_x;Y|\Theta) \label{eqn:baseMI_MITxThetaZero} \\
				&=\Pr \{\Theta=1\} \cdot I(T_x;T_y|\Theta=1), \label{eqn:baseMI_removeNull}
		\end{align}
		
		\vspace{-0.1cm}
		\noindent where \eqref{eqn:baseMI_markov} follows from the Markov chain $T_x \leftrightarrow Y \leftrightarrow \Theta$; \eqref{eqn:baseMI_MITxThetaZero} follows from the fact that the channel input is independent of the selected channel; and \eqref{eqn:baseMI_removeNull} follows from the fact that when $\Theta =0$, no information goes through the channel and therefore $I(T_x;\phi|\Theta=0)=0$.
		Finally, we note that from the definition of $\Theta$, $I(T_x;T_y|\Theta \mspace{-3mu} = \mspace{-3mu} 1) \mspace{-3mu} = \mspace{-3mu} I(T_x;T_y|T_n \mspace{-3mu} \leq \mspace{-3mu} \tau_n)$, and $\Pr \{\Theta \mspace{-3mu} = \mspace{-3mu} 1\} \mspace{-3mu} = \mspace{-3mu} \Pr \{T_n \mspace{-3mu} \leq \mspace{-3mu} \tau_n\} \mspace{-3mu} = \mspace{-3mu} F_{T_n}(\tau_n)$; thus, we obtain \eqref{eq:TCALNchanCap}.
\end{IEEEproof}

\section{Bounds on the Capacity of the DBMT Channel}
\label{sec:capBounds}

Obtaining an exact expression for \eqref{eq:TCALNchanCap} is highly complicated as the maximizing input distribution $f_{T_x}(t_x) \in \mathcal{F}(\tau_x)$ is not known. Therefore, we turn to upper and lower bounds. 
We first note that the conditional mutual information in (\ref{eq:TCALNchanCap}) can be written~as:
\vspace{-0.15cm}
\begin{align}
	I(T_x;T_y| T_n  \leq \tau_n) \mspace{-3mu} &= \mspace{-3mu} h(T_y| T_n \mspace{-3mu} \leq \mspace{-3mu} \tau_n) \mspace{-3mu} - \mspace{-3mu} h(T_y|T_x, T_n \leq \tau_n) \nonumber \\
	& = \mspace{-3mu} h(T_y| T_n \leq  \tau_n) \mspace{-3mu} - \mspace{-3mu} h(T_n| T_n  \leq  \tau_n), \label{eq:Ty_is_a_sum}
\end{align}

\vspace{-0.1cm}
\noindent where \eqref{eq:Ty_is_a_sum} follows from the fact that $T_y = T_x + T_n$ for $T_n \le \tau_n$.
In the following we explicitly evaluate $h(T_n| T_n \leq \tau_n)$ and bound $h(T_y| T_n  \leq \tau_n)$.

\subsection{Characterizing $h(T_n| T_n \leq \tau_n)$}

To characterize the conditional entropy $h(T_n| T_n \leq \tau_n)$ we first define the {\em partial entropy} of a continuous RV $X$, which captures its entropy in the range $(-\infty, \tau]$:
	\begin{definition}[Partial Entropy]
		The partial entropy of a random variable $X$ with PDF $f(x)$ and parameter $\tau \mspace{-3mu} \in \mspace{-3mu} \realSet$, is defined~by:
		\vspace{-0.15cm}
		\begin{align}
		\label{eq:funcN}
			\eta(X,\tau) = - \int_{-\infty}^\tau f(x) \log(f(x)) dx.
		\end{align}
	\end{definition}
	
	\vspace{-0.1cm}
	Let $X$ be a continuous RV with PDF $f_X(x)$ and CDF $F_X(x)$, and let $\tau \mspace{-3mu} \in \mspace{-3mu} \realSet$. The following theorem uses the above definition to characterize $h(X | X < \tau)$:
		\begin{theorem}
			\label{th:entrCondLev}	
			The conditional entropy $h(X|X \le \tau)$ of a continuous RV $X$ is given by:
			\vspace{-0.15cm}
			\begin{align}
			\label{eq:condEntLev}
			h(X|X \le \tau) = \frac{\eta(X,\tau)}{F_X(\tau)} + \log(F_X(\tau)),
			\end{align}
			
			\vspace{-0.1cm}
			\noindent where $\eta(X,\tau)$ is the partial entropy.
		\end{theorem}
		
		\begin{IEEEproof}
			We first note that the RV $\tilde{X}$, defined as $X$ given $X \mspace{-3mu} \le \mspace{-3mu} \tau$, has PDF $f_{\tilde{X}}(\tilde{x}) \mspace{-3mu} = \mspace{-3mu} \frac{f_X(x)}{F_X(\tau)}$. 			
			Next, we write: 
			\vspace{-0.15cm}
			\begin{align}
				\mspace{-5mu} h(\tilde{X}) & = h(X|X<\tau) \nonumber \\
				& \mspace{-5mu} = -\int_{-\infty}^\tau \frac{f_X(x)}{F_X(\tau)} \log\left( \frac{f_X(x)}{F_X(\tau)}\right) dx \label{eqn:timeConstEntropy_1}\\
				& \mspace{-5mu} = \frac{-1}{F_X(\tau)} \int_{-\infty}^\tau \mspace{-2mu} f_X(x) \log(f_X(x)) dx \mspace{-3mu} + \mspace{-3mu}\log(F_X(\tau)) \label{eqn:timeConstEntropy_2} \\
				& \mspace{-5mu} =\frac{\eta(X,\tau)}{F_X(\tau)} + \log(F_X(\tau)), \label{eqn:timeConstEntropy_3}
			\end{align}
			
			\vspace{-0.1cm}
			\noindent where \eqref{eqn:timeConstEntropy_1} follows from the definition of entropy; \eqref{eqn:timeConstEntropy_2} follows by noting that $\int_{-\infty}^{\tau}{f_X(x)dx} = F_X(\tau)$; and \eqref{eqn:timeConstEntropy_3} follows from the definition of $\eta(X,\tau)$.			
		\end{IEEEproof}
		
		As can be seen from Theorem \ref{th:entrCondLev}, to find an expression for the conditional entropy $h(X|X \le \tau)$, for a L\'evy-distributed RV $X$, one needs to find the partial entropy of $X$. This partial entropy is presented in the following lemma:
	\begin{lemma}
		\label{lem:partEntLevy}
		If $X \sim \LevyDist(0,c)$, then
		\vspace{-0.15cm}
		\begin{align}
		\label{eq:funcNlevy}
		\eta(X,\tau) &=\tfrac{1}{2}\log(\tfrac{2 \pi}{c})F_X(\tau)+\tfrac{3}{2}\bigg[(F_X(\tau)-1)\log(\tau)- \nonumber \\
		&  ~~~~4\sqrt{\tfrac{c}{2\pi\tau}}g(c,\tau)\log(e)+\log(c/2)+\gamma\log(e)+2\bigg] \nonumber\\
		&~~~~+ \log(e)\bigg[\tfrac{1}{2}F_X(\tau)+\tau f_X(\tau)\bigg],
		\end{align}
		
		\vspace{-0.1cm}
		\noindent where $f_X(x)$ is given in \eqref{eqn:LevyPDF_0}, $F_X(x)$ is given in \eqref{eqn:LevyCDF}, and $g(c,\tau)$ is a generalized hypergeometric function given by \cite[Ch. 16]{nist10}~as:
		\vspace{-0.15cm}
		\begin{align}
		g(c,\tau) \triangleq~_2F_2(\tfrac{1}{2},\tfrac{1}{2},\tfrac{3}{2},\tfrac{3}{2};\tfrac{-c}{2\tau}).
		\end{align}
	\end{lemma}
	
	\vspace{-0.1cm}
	\begin{IEEEproof}
		The proof is provided in \cite[Appendix B]{journalVersion}.
	\end{IEEEproof}
	
	To find $h(T_n|T_n \le \tau_n)$ we plug \eqref{eqn:LevyPDF_0} and \eqref{eqn:LevyCDF} into \eqref{eq:funcNlevy}, and then plug the resulting expression into \eqref{eqn:timeConstEntropy_3}.
	
	\subsection{Bounds on the Capacity}	
	
	\begin{figure*}[t!]
	\normalsize
	\centering
	\begin{minipage}{.45\textwidth}
		\begin{center}
			\includegraphics[width=0.85\columnwidth,keepaspectratio]{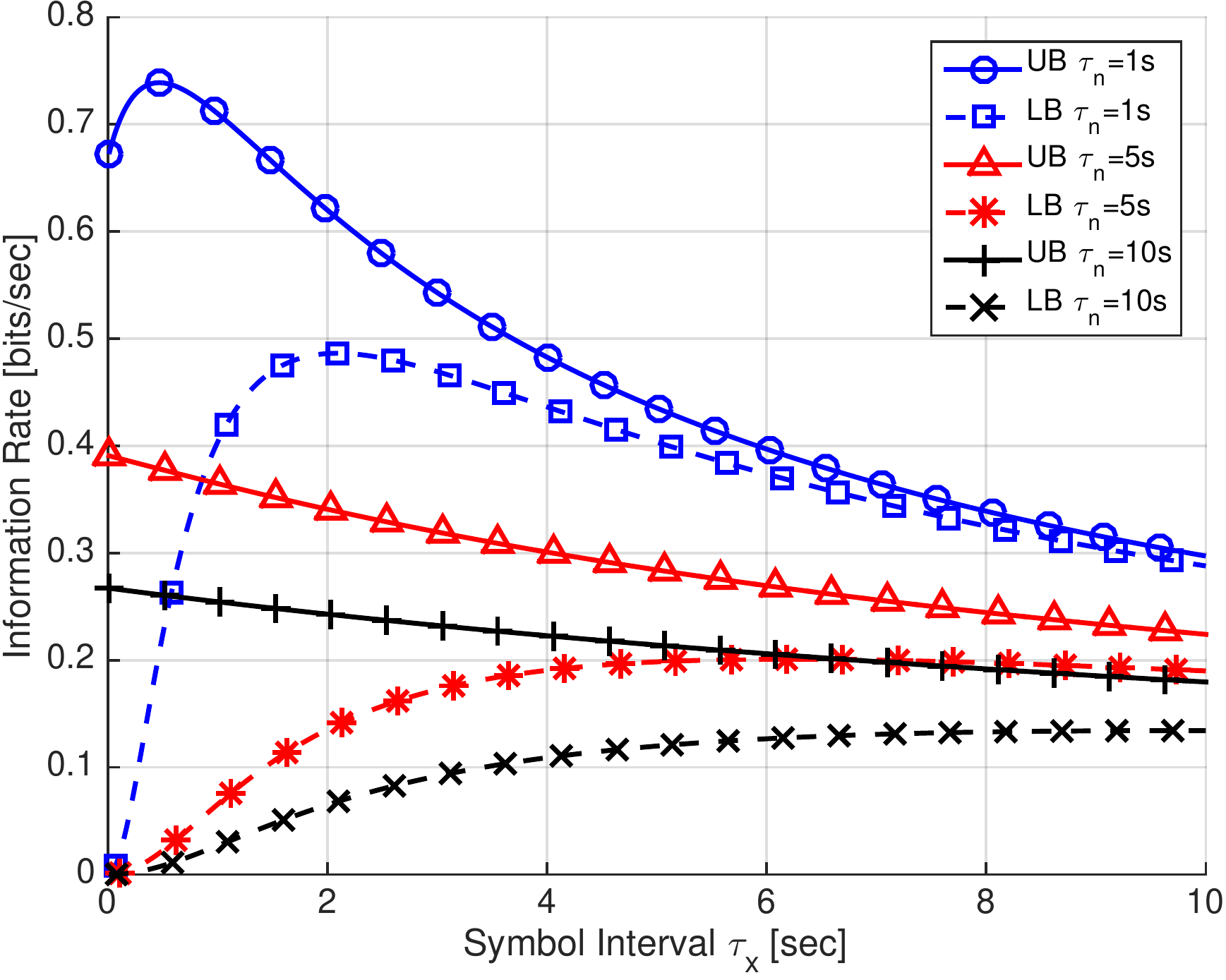}
		\end{center}
		\vspace{-0.35cm}
		\caption{\label{fig:boundsPUC_VStxC0p1} $\mathsf{C}^{\text{lb}}_1(\tau_x, \tau_n)$ and $\mathsf{C}^{\text{ub}}_1(\tau_x, \tau_n)$ versus the symbol interval $\tau_x$, for $\tau_n=1, 5, 10$ [sec], and $c = 0.1$.}
	\end{minipage}
	\vspace{-0.3cm}
	\hspace{0.6cm}
	\begin{minipage}{.45\textwidth}
		\begin{center}
			\includegraphics[width=0.85\columnwidth,keepaspectratio]{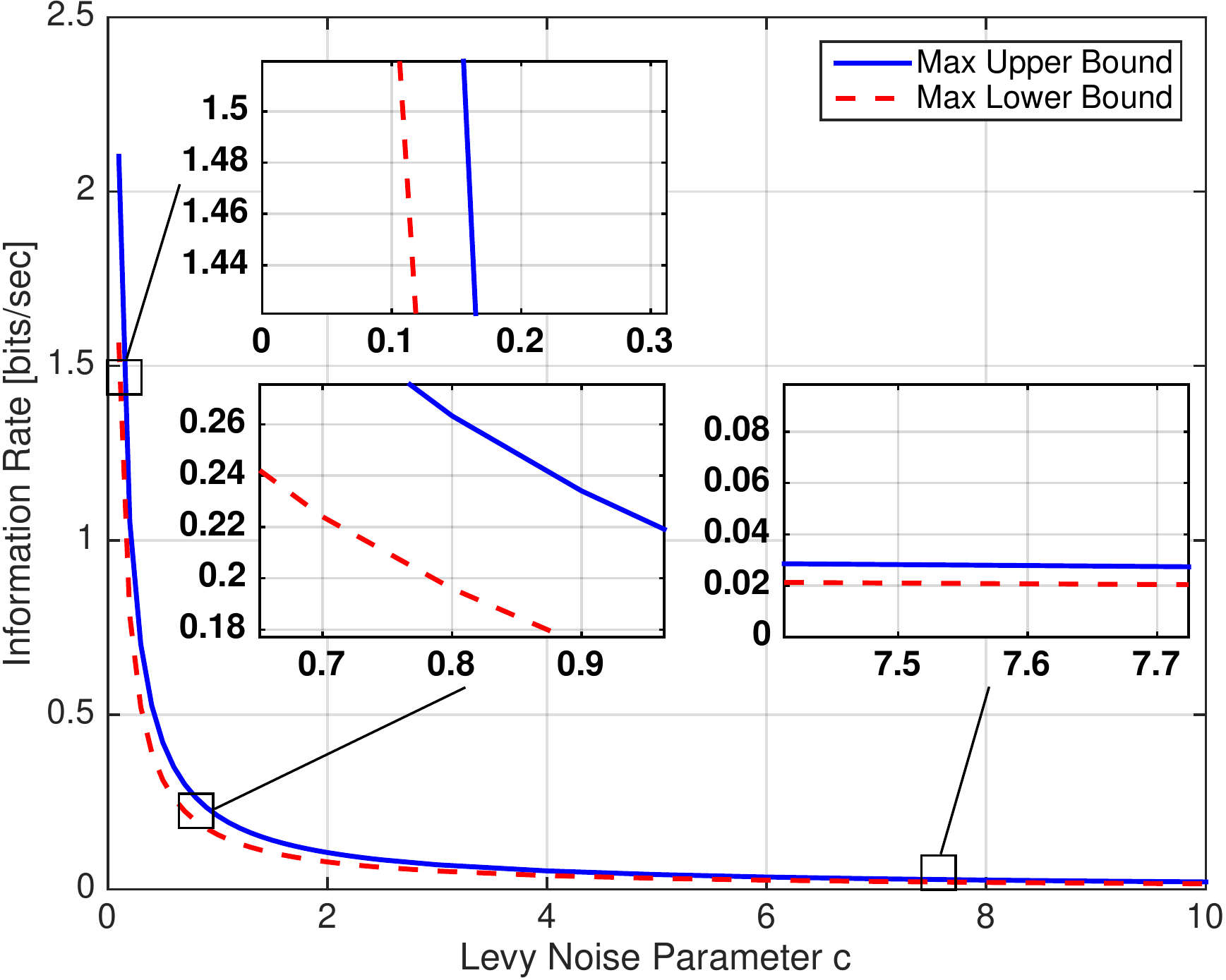}
		\end{center}
		\vspace{-0.35cm}
		\caption{\label{fig:maxBoundsVScn} The maximum lower and upper bounds on the capacity versus the \levy noise parameter $c$. The lower and upper bounds are simultaneously maximized over $\tau_n$ and $\tau_x$.}
	\end{minipage}
	\vspace{-0.3cm}
\end{figure*}
	
Since the maximizing input distribution in \eqref{eq:TCALNchanCap} is not known, it is difficult to obtain an exact expression for the maximal value of $h(T_y | T_n \le \tau_n)$. Therefore, we turn to lower and upper bounds on $h(T_y | T_n \le \tau_n)$, which results in lower and upper bounds on $\mathsf{C}(\tau_n)$. 
For the lower bound we note that $h(T_y | T_n \le \tau_n) = h(T_x + T_n | T_n \le \tau_n)$ and use the entropy power inequality (EPI) \cite[pg. 22]{kim-elgamal-book} to obtain a bound in terms of $h(T_x)$ and $h(T_n | T_n \le \tau_n)$. For the upper bound we again use the relationship $T_y \mspace{-3mu} = \mspace{-3mu} T_x \mspace{-3mu} + \mspace{-3mu} T_n$ to bound $h(T_y | T_n \mspace{-3mu} \le \mspace{-3mu} \tau_n)$ by the logarithm of the support of $T_y$.
Next, define:
\begin{align}
		m(\tau_x,\tau_n,T_n) \triangleq 0.5 \log \big( \tau_x^2+2^{2h(T_n|T_n\leq\tau_n)}\big),
		\label{eqn:mFuncDef}
	\end{align} 

\noindent and recall that $h(T_n|T_n\leq\tau_n)$ is characterized in Theorem \ref{th:entrCondLev}. 
The following theorem presents the lower and upper bounds on $\mathsf{C}(\tau_n)$:
\begin{theorem}
	\label{th:UBLB}
	The capacity of the single-particle DBMT channel is bounded by $\mathsf{C}^{\text{lb}}(\tau_n) \le \mathsf{C}(\tau_n) \le \mathsf{C}^{\text{ub}}(\tau_n)$, where $\mathsf{C}^{\text{lb}}(\tau_n)$ and $\mathsf{C}^{\text{ub}}(\tau_n)$ are given by:
	\begin{align}
	\mspace{-9mu} \mathsf{C}^{\text{lb}}(\tau_n) & \mspace{-1mu} \triangleq \mspace{-1mu} \underset{\tau_x}{\max} \frac{\left(m(\tau_x,\tau_n,T_n) \mspace{-4mu} - \mspace{-4mu} h(T_n|T_n \mspace{-3mu} \leq \mspace{-3mu} \tau_n)\right) \mspace{-2mu} F_{T_n} \mspace{-2mu} (\tau_n)}{\tau_x + \tau_n} \label{eq:capacityLB_single} \\
	\mspace{-9mu} \mathsf{C}^{\text{ub}}(\tau_n) & \mspace{-1mu} \triangleq \mspace{-1mu} \underset{\tau_x}{\max} \frac{\left(\log(\tau_x \mspace{-3mu} + \mspace{-3mu} \tau_n) \mspace{-3mu} - \mspace{-3mu} h(T_n|T_n \mspace{-3mu} \leq \mspace{-3mu} \tau_n)\right) \mspace{-2mu} F_{T_n} \mspace{-2mu} (\tau_n)}{\tau_x+\tau_n}. \label{eq:capacityUB_single}
	\end{align}
\end{theorem}

\begin{IEEEproof}
	For the lower bound $\mathsf{C}^{\text{lb}}(\tau_n)$ we write:
	\begin{align}
		h(T_y| T_n \leq \tau_n) &= h(T_x+T_n| T_n \leq \tau_n) \nonumber \\
											&\geq 0.5 \log \bigg( 2^{2h(T_x)}+2^{2h(T_n|T_n \leq \tau_n)}\bigg), \label{eq:powerInq_2}
	\end{align}
	where \eqref{eq:powerInq_2} follows from the EPI, and also by noting that $T_x$ and $T_n$ are independent given $T_n \leq \tau_n$.
	Furthermore, as this bound holds for every $f_{T_x}(t_x)$, we use the entropy maximizing distribution for $T_x$, the uniform distribution, with entropy $\log(\tau_x)$ to obtain $m(\tau_x,\tau_n,T_n)$.
	
	For the upper bound $\mathsf{C}^{\text{ub}}(\tau_n)$ we write:
		\begin{align}
		h(T_y| T_n \leq \tau_n) &\leq \log(\tau_x + \tau_n), \label{eqn:capacityUB} 
		\end{align}
		
		\noindent where \eqref{eqn:capacityUB} follows since the uniform distribution maximizes entropy over a finite interval.
\end{IEEEproof}

In the next section we will numerically evaluate the bounds on the capacity of the DBMT channel derived in this section, and show they are tight for large values of the symbol interval.
\vspace{-0.1cm}
\section{Numerical Results} \label{sec:Numerical}
\vspace{-0.1cm}

Fig. \ref{fig:boundsPUC_VStxC0p1} depicts the effect of symbol interval on channel capacity by plotting the bounds on capacity versus $\tau_x$, for $\tau_n=1,5,10$ [sec], and for  $c = 0.1$. As the values of $\tau_x$ tend to zero, the bounds are not tight, while as $\tau_x$ increases they converge. For smaller values of particle's lifetime $\tau_n$, the bounds tend to converge more rapidly. 
Note that in Fig. \ref{fig:boundsPUC_VStxC0p1}, for a given $\tau_n$, the lower and upper bounds are maximized by different values of $\tau_x$. Therefore, it is not clear from the plots which value of $\tau_x$ maximizes the capacity. However, it can be observed that the bounds achieve their maximal values for relatively small values of $\tau_x$. 

Next, we study the effect of the \levy noise parameter $c$ on the capacity of the DBMT channel. Here, we numerically maximize the lower and upper bounds on the capacity with respect to {\em both} $\tau_x$ and $\tau_n$. 
Fig.~\ref{fig:maxBoundsVScn} depicts the maximal lower and upper bounds as a function of $c$. 
It can be observed that the capacity drops very rapidly with respect to $c$ and that the bounds are relatively tight. We note here that the increase in $c$ can result from either an increase in the distance between the transmitter and the receiver, or a decrease in the diffusion coefficient of the information particles with respect to the propagation medium. For instance, the diffusion coefficient for glucose in water at 25$^\circ$C is 600 $\mu$m$^2$/s \cite{far16ST}. Thus, if $r \mspace{-3mu} = \mspace{-3mu} 10\mu$m, the \levy noise parameter is $c=0.083$, and if $ r \mspace{-3mu} = \mspace{-3mu} 50\mu$m, then $c= 2.083$. If glycerol is used instead of glucose, then $d \mspace{-3mu} = \mspace{-3mu} 930 \mu$m$^2$/s \cite{far16ST}, and the \levy noise parameters would be $c=0.054$ and $c=1.344$, respectively.

\vspace{-0.1cm}
\section{Conclusions} \label{sec:concl}

\vspace{-0.1cm}
In this work we considered MT channels, where information is modulated on the release time of particles, and showed that these channels can be represented as additive noise channels. By assuming that the information particles have a finite lifetime, we formally defined the capacity of the MT channels. We then showed that the \levy distribution can be used to formulate the DBMT channel, and derived upper and lower bounds on capacity of this channel. 
Finally, we numerically evaluated these bounds and numerically showed that the bounds converge when the symbol interval is large (also analytically shown in \cite[Remark 5]{journalVersion}).

\bibliographystyle{IEEEtran}
\vspace{-0.1cm}
\bibliography{IEEEabrv,MolCom}

\end{document}